\documentclass[aps,prl,groupedaddress,showpacs,preprintnumbers,amsmath,amssymb,twocolumn]{revtex4}

\usepackage{graphicx}
\usepackage{dcolumn}
\usepackage{bm}

\begin{document}

\title{Statistics of Particle Trajectories at Short Time Intervals Reveal fN-Scale Colloidal Forces}

\author{Sunil K. Sainis}
\author{Vincent Germain}
\author{Eric R. Dufresne}
 \affiliation{Departments of Mechanical
 Engineering, Chemical Engineering and Physics,
 Yale University, New Haven, CT 06511.}
 \email{eric.dufresne@yale.edu}

\date{\today}

\begin{abstract}

We describe and implement a technique for extracting forces from the
relaxation of an overdamped thermal system with normal modes.  At
sufficiently short time intervals, the evolution of a normal mode is
well described by a one-dimensional Smoluchowski equation with
constant drift velocity, $v$, and diffusion coefficient, $D$. By
virtue of fluctuation-dissipation, these transport coefficients are
simply related to conservative forces, $F$, acting on the normal
mode: $F = k_BT v/D$. This relationship implicitly accounts for
hydrodynamic interactions, requires no mechanical calibration, makes
no assumptions about the form of conservative forces, and requires
no prior knowledge of material properties. We apply this method to
measure the electrostatic interactions of polymer microspheres
suspended in nonpolar microemulsions.

\end{abstract}


\maketitle

The structure and stability of colloidal dispersions depend
sensitively on the interactions of suspended particles.  An early
triumph of colloid science was the theory of Derjaguin-Landau-Verwey
and Overbeek (DLVO) \cite{derjaguin.1941, verwey.1948} which
juxtaposes short-range Van der Waals forces and longer-range
electrostatic forces to characterize the stability of aqueous
colloidal dispersions.
%
Yet the role of electrostatic interactions in nonpolar solvents has
remained controversial \cite{morrison.1993}.  Electrostatic forces
between surfaces in nonpolar solvents have recently been reported
for a variety of surfaces in nonpolar microemulsions
\cite{briscoe.2002,mcnamee.2004,hsu.2005}. In certain regimes,
measured interactions \cite{hsu.2005} are identical to the
screened-Coulomb component of the standard DLVO theory. In others, a
novel counterion-only double-layer theory is needed to describe
observed forces \cite{attard.2002}. While the reality of
electrostatic interactions in nonpolar environments has been
established, their origin and significance remain mysterious. To
that end, robust methods for measuring interparticle forces are
needed to bring out the underlying physics.

A variety of methods have emerged to directly measure colloidal
forces.  The surface forces apparatus
\cite{israelachvili.1992,briscoe.2002} and atomic force microscope
\cite{ducker.1991,mcnamee.2004} measure forces mechanically and have
respective force resolutions at
 the nN and pN scales. Alternatively, native thermal fluctuations can reveal interparticle forces.
   Such
methods are well suited to real-space imaging and provide force
resolutions on the fN scale. For weakly interacting systems, the
potential can be extracted from the equilibrium distribution
\cite{prieve.1990, biancaniello.2006} by inverting the Boltzmann
equation
 $U/k_BT = -\ln P_{eq}$.
Similarly, liquid-structure theory or reverse Monte Carlo methods
enable the extraction of pair-potentials from the pair-correlation
function, $g(r)$, of stable semi-dilute dispersions of identical
particles \cite{kepler.1994,behrens.2001,hsu.2005}. This method
assumes pair-wise additivity of potentials \cite{brunner.2002} and
is difficult to implement without introducing artifacts from
confining surfaces.  All of these equilibrium methods are limited to
interactions of less than a few $k_BT$.  An  alternative approach,
due to Crocker and Grier \cite{crocker.1994}, analyzes the dynamics
of a system relaxing toward equilibrium. Their method, Markovian
Dynamics Extrapolation (MDE), offers the attractive advantage of
sampling higher interaction energies by driving the system out of
equilibrium with an external force.  MDE elegantly identifies the
equilibrium distribution, $P_{eq}$, as an eigenvector (with
eigenvalue one) of the experimentally sampled finite-time
propagator. However, forces are not extracted from local properties
of the trajectories. Rather, the propagator must be thoroughly
sampled over the full range of the interaction - from hard-core
repulsion at short-range to zero force at long-range.
 Furthermore, systematic effects due to sampling errors on the calculation of the
eigenvectors are hard to quantify and artifacts from hydrodynamics
are difficult to rule out.

In this Letter, we present a simple method for extracting
conservative forces between isolated pairs of colloidal particles
from the statistics of their trajectories at short time intervals.
While our experimental apparatus is a straight-forward extension of
the blinking optical tweezers introduced by Crocker and Grier
\cite{crocker.1994}, we propose an alternative method of data
analysis that measures forces locally and implicitly accounts for
hydrodynamic coupling.  We apply this method to characterize the
electrostatic interactions of polymer colloids suspended in a
nonpolar microemulsion.

We measure the electrostatic interactions of carboxylate modified
polystyrene latex particles, radius $a=600$ nm (Interfacial Dynamics
Corp.), suspended at vanishingly small volume fraction, $\phi \le
10^{-6}$, in a nonpolar microemulsion of AOT (sodium
di-2-ethylhexylsulfosuccinate) in hexadecane.  Samples are prepared
and stored in a low-humidity glove box. A glass chamber, constructed
from a standard microscope slide, three microscope coverslips (No
1.5), and UV curing epoxy (Norland 61), holds the sample for optical
microscopy and micromanipulation. An inverted optical microscope
(Nikon TE2000) equipped with an oil immersion lens (100X, N.A. 1.4)
images the suspension in bright field. Images are recorded on a high
speed digital video camera (Photron Fastcam 1024PCI) at a frame rate
of 500Hz.  Centroid algorithms \cite{crocker.1996a}, implemented in
MATLAB, locate particle centers to a resolution of about 10 nm.

\begin{figure}\includegraphics[scale=0.8]{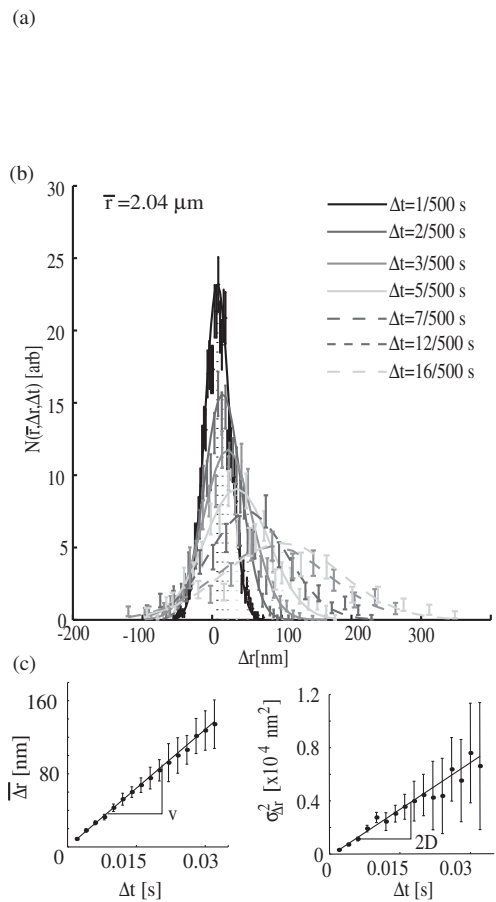} \caption{
\label{fig:mov} Statistics of Particle Trajectories at Short Time
Intervals (a) Typical images of trapped and free microspheres in 1
mM AOT/hexadecane. (b) A histogram of $\Delta r$ at various $\Delta
t$ is plotted for an average separation $\bar r=$ 2.04 $\mu$m. (c)
Fits of the mean displacement and the mean squared-displacement as a
function of time yield two transport coefficients $v$ and $D$,
respectively.}\end{figure}

We extract interparticle forces from the statistical properties of
the trajectories of isolated pairs of beads.  We use blinking
optical tweezers \cite{crocker.1994} to repeatedly trap and release
particles at a desired separation, as shown in Figure
\ref{fig:mov}(a).
 This method allows us to efficiently
acquire good statistics for the trajectories of spheres in unlikely
configurations. In our setup, described elsewhere
\cite{chapin.2006}, a pair of optical traps is made with the 532nm
output of a diode-pumped solid state laser (Coherent Verdi V-5)
using holographic optical tweezers \cite{dufresne.1998,
dufresne.2001, liesener.2000}. To avoid wall effects
\cite{larsen.1997, dufresne.2000, squires.2000}, particles are
trapped at least 10$\rm{\mu}$m from the walls of the sample chamber.
Once the optical tweezers have set the height and initial separation
of the beads,
 we blink the laser using a chopper (Thorlabs MC1000A) at a
rate $1/\tau=20$ Hz, with a duty cycle of 1:6.  While the laser is
off, the particles move freely, traveling distances up to about 200
nm. This motion is a combination of thermal diffusion and drift
induced by interparticle forces.  As the solvent does not absorb a
measurable amount of trapping light, we expect that local heating by
the optical tweezers is not significant. Furthermore, if any heat
were delivered by the laser, it would  be dissipated within
microseconds by thermal conduction.  By studying the motions of the
particles only when the trap is off, we ensure that our measurements
are insensitive to the details of the interaction of the trap with
the particles and to any interactions between particles due to light
scattering in the traps \cite{burns.1989}.

We characterize the stochastic trajectories of free particles with
time-dependent two-particle probability distributions. We reduce the
trajectories to a list of statistically independent events,
characterized by initial and final particle separations, $r_i$ and
$r_f$, and a time interval, $\Delta t$. Each event is assigned to a
spatial bin according to its average separation, $\bar r =
(r_i+r_f)/2$. We count the number of events with a particular value
of displacement, $\Delta r=(r_f-r_i)$, for each separation, $\bar
r$, and time interval, $\Delta t$. While each event describes
particle dynamics between a single pair of laser flashes, we
concatenate events across many flashes to get good statistics. The
resulting histogram, $N(\bar r,\Delta r, \Delta t)$, represents the
time-dependent two-particle probability distribution, and is
well-fit by a Gaussian curve. Histograms for different values of
$\Delta t$ at an average separation $\bar r=$2.04 $\mu$m are plotted
in Figure \ref{fig:mov}(b).
 The mean
displacement, $\overline { \Delta r} $, and  the variance of the
displacement, $\sigma^2_{\Delta r}$, increase linearly with time, as
shown in Figure \ref{fig:mov}(c). The slope of the mean displacement
provides a drift velocity, $v$. Likewise, the slope of the variance
provides a diffusion coefficient, $D$. We observe these linear
relationships at all particle separations.

\begin{figure}
\includegraphics[scale=0.8]{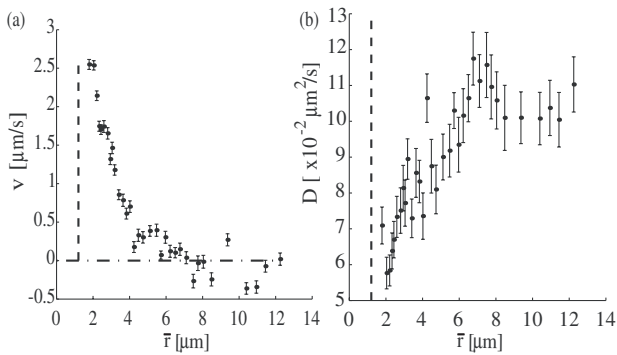} \caption{ \label{fig:vd} Separation dependence
of (a) $v(r)$, the mean velocity of separation, and (b) $D(r)$, the
relative diffusion coefficient in 1 mM AOT/hexadecane.
}\end{figure}

The separation dependence of the velocity and diffusion coefficients
are plotted in Figure \ref{fig:vd}.
 The relative velocity is positive and decreases slowly
from a maximum at the smallest separation, suggesting a long-range
repulsive force. In contrast, the diffusion is suppressed by viscous
forces as particles come near contact.  At low Reynolds number,
force and velocity are related through the hydrodynamic mobility
tensor, $\vec v=\mathbf{b} \cdot \vec F$. The mobility,
$\mathbf{b}$, generally depends on the size and separation of the
spheres and the viscosity of the solvent. If the particles and
solvent have been characterized in separate experiments, then the
mobility  can be calculated from existing theory
\cite{batchelor.1976}. However, typical expressions for hydrodynamic
mobility are complicated by the presence of surface charge and
counterions \cite{obrien.1978}. Without \emph{a priori} knowledge of
the zeta potential and screening length, these challenging
calculations become intractable. We sidestep this obstacle by
exploiting our simultaneous measurement of the relative velocity and
diffusion coefficients.

To clarify the relationship between the transport coefficients and
 forces, let us consider a generalized form of Fick's
first law for a system of $N$ interacting particles
\cite{Dhont.1996}:
\begin{equation} \label{eq:th1}
\vec{J}_{i}=\sum_j \left(-\mathbf{D}_{ij} \cdot [\vec{\nabla}_{j} P
+ \beta (\vec{\nabla}_{j} U) P] \right ).
\end{equation}
Here, $\vec{J}_{i}$ is the probability current of the $i^{\rm{th}}$
particle, $\mathbf{D}_{ij}$ is the diffusivity tensor,
$P=P(\vec{r}_{1},\ldots \vec{r}_{N},t)$ is the probability
distribution, $\beta=1/k_BT$, and $U(\vec{r}_{1},\ldots\vec{r}_{N})$
is the
 potential energy.  The
first term in Equation (\ref{eq:th1}) is the current due to coupled
diffusion and the second term captures drift due to conservative
 forces. Enforcing the conservation of probability, we
arrive at the Smoluchowski equation:
\begin{equation} \label{eq:th2}
\partial_{t}P=\sum_i -\vec{\nabla}_{i} \cdot \vec{J}_{i}=
  \sum_{i,j} \left( \vec{\nabla}_{i} \cdot \mathbf{D}_{ij} \cdot [\vec{\nabla}_{j} P + \beta
(\vec{\nabla}_{j} U) P] \right).
\end{equation}
The Onsager relations \cite{pathria.1996} demand that,
$\mathbf{D}_{ij}=\mathbf{D}_{ji}$. Therefore, $\mathbf{D}_{ij}$ can
be diagonalized by a set of normal coordinates, $\vec x_1 \ldots
\vec x_N$. If the interaction potential is a linear combination of
contributions from each mode, $U=\sum_i U_i(\vec x_i)$, then the
probability distribution can be separated, $P=\prod_i P_i(\vec
x_i)$, and
\begin{equation} \label{eq:th2b}
\partial_{t}P_i= \vec{\nabla}_{i} \cdot [ \mathbf{D}_{i} \cdot \vec{\nabla}_{i} P_i
- \vec{v}_i P_i],
\end{equation}
where $\vec{v_i}=-\beta \mathbf{D}_i \cdot \vec{\nabla}_i U_i$. Here
we have restricted our analysis to short time intervals so that the
spatial dependence of the normal modes can be ignored. This is
valid, provided that: $\overline{\Delta x_i} v_i' \ll v_i$,
$\sigma_{\Delta x_i} v_i' \ll v_i$, $\overline{\Delta x_i} D_i' \ll
D_i$, and $\sigma_{\Delta x_i} D_i' \ll D_i$, where primed variables
indicate spatial derivatives.  Furthermore, the vector nature of the
normal modes can be ignored provided that their displacements are
small compared to their magnitude: $\overline{\Delta x_i} \ll x_i$
and $\sigma_{\Delta x_i} \ll x_i$. This leads to a tractable
one-dimensional form for the Smoluchowski equation:
\begin{equation}\label{eq:th19}
\dot{P_i} = D_i P_i'' - (v_i - D_i') P_i'.
\end{equation}
For $D_i' \ll v_i$, this is solved by a Gaussian distribution,
\begin{equation}\label{eq:th15}
P(\Delta x_i,\Delta t)=(2 \pi \sigma^2_{\Delta x_i} )^{-1/2}
 \exp \left[
-\frac{ \left( \Delta x_i- \overline{ \Delta x_i} \right) ^2}{2
\sigma^2_{\Delta x_i}} \right],
\end{equation}
where $\overline{\Delta x_i } = v_i \Delta t,$ and $\sigma^2_{\Delta
x_i} =2 D_i \Delta t$. Thus, we arrive at a convenient expression
for the conservative force acting on the $i^{\rm{th}}$ normal mode,
\begin{equation}\label{eq:fvr2} F_i=-U_i'=k_{B}T \frac{v_i}{D_i}.
\end{equation}
This equation, a direct consequence of fluctuation-dissipation,
relates the force to locally measured transport coefficients without
appealing to particular models of hydrodynamic or conservative
forces.

\begin{figure}\includegraphics[scale=0.8]{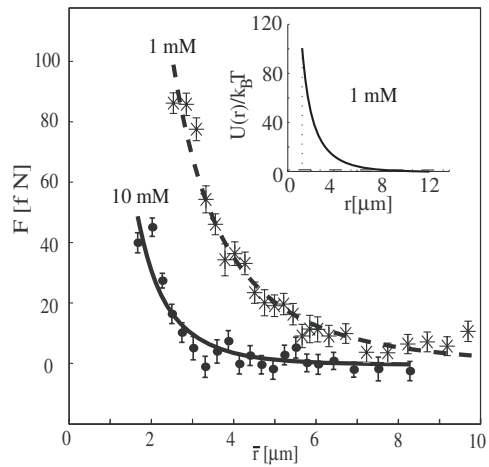}
\caption{ \label{fig:forces} Electrostatic forces between charged
colloidal particles in a nonpolar solvent. Repulsion between
carboxylate modifed PS at two concentrations of AOT, as noted.
Curves are fits to screened-Coulomb interactions \emph{(inset)}
Interaction energy, $U(r)/k_BT$, implied by a fit at 1 mM
AOT.}\end{figure}

This analysis is readily applied to identical colloidal particles,
where the separation vector, $\vec r$, is a normal mode of the
diffusion tensor and $U=U(\vec r)$.  The interparticle forces can be
directly calculated by dividing the drift velocities by the
diffusivities at each separation found in Figure \ref{fig:vd}. The
resulting force profile is shown in Figure \ref{fig:forces}.
 These spheres show purely repulsive interactions
with a maximum measurable force of about 100 fN and a resolution of
a few fN.  The interparticle forces fall off slowly as the particle
separation increases, with measurable repulsions out to about five
particle diameters. These results are well fit by a screened-Coulomb
form,
\begin{equation} \label{eq:dlvof}
F(r)=k_{B}T\left(\frac{e\zeta}{k_BT}\right)^2\frac{a^2}{\lambda_B}\frac{e^{-\kappa(r-2a)}}{r}
\left( \frac{1}{r}+\kappa \right),
\end{equation} where the Bjerrum length $\lambda_B=e^2 /4 \pi \epsilon \epsilon_o k_B
T$. This fit returns a screening length $\kappa^{-1}=5.0 \pm 0.2$
$\mu$m and an apparent surface potential, $|e\zeta/k_BT|=3.30 \pm
0.04$. It is important to note that the fitted value of $|e\zeta|$
reflects the surface potential as seen from long range. This value
will be smaller than the actual surface potential for highly charged
surfaces due to non-linear screening near the particle surface.

Interparticle forces vary with the concentration of surfactant. When
the concentration of AOT is increased to 10 mM, the range and scale
of the interparticle forces drop, as shown in Figure
\ref{fig:forces}. In particular, the screening length is lowered to
$0.6 \pm 0.1$ $\mu$m and the apparent surface potential is
significantly reduced $|e\zeta/k_BT|=1.8 \pm 0.1$.  Previous work on
a related system \cite{hsu.2005} reported similar values of $\kappa$
and $\zeta$.

Interaction potentials, $U(r)$, can be calculated from these
parameters, as demonstrated in the inset of Figure \ref{fig:forces}.
The potential is soft and long-ranged, decaying from about $100$
$k_BT$ at contact to $<k_BT$ at about $r = 10$ $\rm{\mu}$m. We use
fitted values of $\kappa^{-1}$ and $|e\zeta/k_BT|$ because we have
found direct integration of the force curve to be highly unreliable;
by simply varying the size of the spatial bins, conventional
potentials with screened-Coulomb forms can be transformed
 into anomalous potentials with long-range attractions.

We present a method for extracting the conservative forces between
colloidal particles from the statistics of their trajectories.  This
method requires no separate measurements of solvent and particle
properties.  The only supporting measurements are the spatial and
temporal calibrations of the imaging system and the temperature of
the sample. Additionally, our measurement is completely independent
of specific models for hydrodynamic and electrostatic interactions
between particles.  Our method of data analysis, rooted in the
general principles of nonequilibrium statistical mechanics, may be
extended to probe generalized forces acting on fluctuating normal
modes of any thermal system, from more complex colloidal systems to
the internal dynamics of molecules.

\begin{acknowledgments}
We thank Ian Morrison and Todd Squires for helpful discussions and
Cabot Corporation for support.
\end{acknowledgments}


\end{document}